\documentclass{article}
\usepackage[utf8]{inputenc}
\usepackage{url}
\usepackage{color}
\usepackage{graphicx}
\usepackage{bm, amsmath}
\usepackage{hyperref}
\usepackage{subcaption}

\title{Parameter estimation of epidemic spread in two-layer random graphs by classical and machine learning methods}
\author{Ágnes Backhausz\footnote{Corresponding author. ELTE Eötvös Loránd University, Budapest, Hungary and HUN-REN Alfréd Rényi Institute of Mathematics. Email: \tt{agnes.backhausz@ttk.elte.hu}}, Edit Bognár, Villő Csiszár, \\ Damján Tárkányi, András Zempléni\footnote{ELTE Eötvös Loránd University, Budapest, Hungary, Faculty of Science, Department of Probability Theory and Statistics.}} 
\date{3 July 2024}

\begin{document}
\maketitle

\begin{abstract} Our main goal in this paper is to quantitatively compare the performance of classical methods to XGBoost and convolutional neural networks in a parameter estimation problem for epidemic spread. As we use flexible two-layer random graphs as the underlying network, we can also study how much the structure of the graphs in the training set and the test set can differ 
while to get a reasonably good estimate. In addition, we also examine whether additional information (such as the average degree of infected vertices) can help improving the results, compared to the case when we only know the time series consisting of the number of susceptible and infected individuals.  Our simulation results also show which methods are most accurate in the different phases of the epidemic.

\end{abstract}

\section{Introduction}

As the SARS-CoV-2 pandemic starting in 2020 and the current potential risk of the avian influenza HPAI A(H5N1) epidemic shows, it is often crucial to identify the emergence of new infectious diseases as soon as possible and to perform effective preventive actions. From a mathematical point of view, among other aspects, estimating the infection rate (the probability that an infected individual transmits the disease to one of its contacts in a given time period) precisely plays a key role, as this serves as the base of the prediction of the spread of the epidemic, and hence this can be used to choose the optimal strategies of prevention. The starting point of  this statistical problem is the following question: which types of data are essential to get a good estimate, and whether there are other observations that contain further information and can be used to improve the quality of the estimates. Once we have the data, we still have several methods that we can use to get good estimates: a variety of methods are available, including classical tools and methods from statistical machine learning. Our goal in this paper is to compare estimates provided by a classical method, the XGBoost and neural networks, in different periods of the epidemic spread process, by modifying also the set of characteristics that are used. These quantitative comparisons will be useful in understanding the advantages and drawbacks of the different methods, and on the possibilities of improving the estimates by collecting more data.

Among the various ways of modelling epidemic spread, we will use two-layer random graphs with different structure.  The 
different layers 
represent different types of communities \cite{fourlayer, britton_oneill, britton, multilevel}; see also the recent work \cite{kubaschmulti} on a limit approach of these multilayer graphs. As proposed in the household
model introduced and analysed by Ball \cite{household, config_sir}, we start with a layer for households: this consists of 
disjoint, small complete graphs. Next we add  a layer representing the connections between households, which is in our case a random graph, the
{\it polynomial graph model} defined by Ostroumova, Ryabchenko and Samosvat \cite{clustering}. In particular, we utilise the 
three-parameter model described in Section 5.1 of the above paper.
We compare the results for this graph with scale-free property, to another type of graphs, where the second layer is also homogenous.

Our aim is to investigate the properties of the modern machine learning approaches for estimating the parameter of interest in this complex model. We use XGBoost, see e.g. \cite{xgb}, and compare its performance to other approaches like convolutional neural networks. We also refer to our recent paper \cite{elsocikk}, where we studied a similar problem, using
a specific graph neural network 
in a case when much more information is available from the underlying network than in the current case. 

Other approaches for using machine learning methods (mostly graph neural networks) for the solution of problems related to epidemic spread are also available, here we only present briefly a few of these. We first refer to \cite{tomy2022estimating}. In this work, Tomy et al.\ use graph neural networks to obtain as detailed information as possible on the actual state of the individuals in the network, that is, to get more information than we can achieve with the usual testing capacity. Song et al.\ \cite{song2023covid} predict the probability that the $n$th infected individual transmits the disease, by using both graph neural networks and graph attention networks. Kapoor et al.\cite{kapoor2020examining} use real data with spatio-temporal mobility datasets (that is, graphs in which one can model connections between vertices as a function of space and time) to forecast the spread of the epidemics, by taking into account the interactions between different regions of the US. Their method is also based on graph neural networks, and can be used widely, as it is not relying on strict assumptions on the spreading process of the graph. Shah et al.\ \cite{shah2020finding} propose machine learning algorithms to find patient zero. In \cite{mevznar2021prediction}, Me\v znar et al.\ study the predictability of the epidemic spread process by using graph neural networks, and focusing on the role of the individual from whom the infection process starts. Compared to these works, our paper focuses on a simple random graph model with tunable parameters, where the effect of the structure of the graph can also be studied, and we compare the  
neural network methods with classical and XGBoost models,
 providing a more detailed understanding of the effectivity of these methods for estimating the infection parameter. 

In Section 2, we briefly present the random graph models that we studied.  Section 3 is about the  
methods that we used for estimating the infection parameter. Section 4 contains the results of the computer simulations and the comparison of the different methods, also examining the effect of the structure of the training set. Section 5 contains our main conclusions.

\subsection*{Acknowledgement.} This research was supported by the National Research, Development and Innovation Office within the framework of the Thematic Excellence Program 2021 - National Research Sub programme: “Artificial intelligence, large networks, data security: mathematical foundation and applications”.

\section{The graph models}

One of our main goals is to understand the effect of the graph structure on the efficiency of the estimating algorithms. In order to include the community structure of real-world networks, our starting point is the household model of Ball (see e.g.\ \cite{household, config_sir}). In the original, perfectly homogeneous model, members of different households are all connected to each other with smaller weight. Instead of this complete graph, we use more complex and more flexible random graphs as a second layer, connecting the households.  To examine the effect of the heterogeneity in the degrees, we used two different graph models: one with a scale-free property, where the degree distribution follows a power law,
and another one with a homogeneous degree distribution.

The first layer of the graph is identical in both models.
For sake of simplicity, let us fix the size of the households, $N_{\rm hh}=5$. Given the total number of vertices, $n$, we form disjoint cliques (complete graphs) of size $N_{\rm hh}$ from these vertices, so that less than $N_{\rm hh}$ vertices are left out. These cliques can represent other groups of people who meet regularly, almost every day: schoolmates, colleages etc., this is why we have chosen the size of the cliques to be larger than the typical household size.  Still, we will refer to this as the household layer of the graph. 

To represent the different role of connections within and between households, the edges are weighted. In particular,   
every edge in the household layer  has weight $1$, while we denote by $w$ the weight of the edges in the second layer. 

\subsection{Scale-free second layer}
\label{sec:pamodel}

The first model that we used is a combination of the household layer with the polynomial model of Ostroumova, Ryabchenko and Samosvat \cite{clustering} as the second layer. Here we only give the main elements, a more detailed description can be found in \cite{elsocikk}. 

Once we have formed the first layer, we overlay a second layer on it,  representing  pairs of individuals who meet regularly, but possibly less frequently. This random graph is chosen independently of the cliques in the household layer. In each step, a vertex with $m$  new edges is added to the graph. The flexibility of the model is guaranteed by the parameters $p_{\rm pa}$, $p_{\rm u}$ and $p_{\rm tr}$ (denoted by $\alpha, \delta, \beta$, respectively, in the original paper), which are the probabilities of applying the preferential attachment rule, or the uniform choice, or adding triangles, respectively, when we form the edges of the new vertex. The initial graph is an arbitrary graph with
$n_0$ vertices and $mn_0$  
edges. It is shown in \cite{clustering} that this graph is scale-free if $p_{\rm u} < 1$, with the power-law exponent $\gamma = 1 + \frac{2}{2p_{pr} + p_{tr}}$ (Ostroumova, Section 5.2). In the sequel, we will use $m=4$, which means that the average degree of a vertex outside its household is equal to $8$. 

\subsection{Cliques as second layer}

\label{sec:modeltwo}

Real-world social networks often have an important structural property which cannot be described easily by preferential attachment dynamics or adding triangles, as in the model described in the previous subsection. In particular, most people typically belong to more than one clique: a clique corresponding to the household where this person lives, and a clique corresponding to a workplace or school. As the work \cite{bokanyi2022anatomy}  shows, in some situations this might also be a better graph model to match the real observations, as certain authorities collect data in this form, by registering the members of households, schools and workplaces.  

To construct such a model, once we have formed the first layer of the graph, 
we fix the size of workplaces (which can of course represent various other groups of people meeting each other every day).   Then, independently of the layer of households, we form disjoint cliques of size $N_{\rm wp}$ randomly. 

For the modeling of workplaces in the second layer, a modified version usually referred to as the relaxed caveman graph  \cite{wattsnetworks} was also examined. In the relaxed caveman graph, after generating isolated cliques of size $N_{\rm wp}$, each edge $uv$ is revisited with a given probability $p_{\rm relaxed}$ and rewired to $u \tilde{v}$, where node $\tilde{v}$ is drawn uniformly. If edge $u \tilde{v}$ already exists, no rewiring takes place. With this relaxation, the given graph still contains fairly well-defined communities, but randomness is also taken into consideration.

In the basic scenario, we set $N_{\rm wp}=9$, because with this choice the degree of a vertex outside its household will be equal to $8$, which is the average degree outside the household in our simulations for the household graph with scale-free second layer (recall Subsection \ref{sec:pamodel}).

\section{Simulation of the epidemic spread process and estimating the infection parameter}

Once we have the underlying random graph (either according to Subsection \ref{sec:pamodel} or to Subsection \ref{sec:modeltwo}), the next step is to simulate the epidemic spread. For this, we have chosen one of the simplest models, the well-known SIR model, with susceptible (S), infected (I) and recovered (R) vertices (see e.g.\ \cite{simon}). In the beginning, a randomly chosen 
low percentage of the vertices were infected. 
For the computer simulations, we used the Gillespie algorithm, in the form as it is presented in \cite{simon}. The recovery rate was fixed $\gamma=1$. The infection rate between vertices connected in the first layer is $\tau$, and it is $\tau \cdot w$ on edges belonging to the second layer. In our setup, we assume that connections in the second layer are weaker, this is represented by $0<w<1$. In our simulations, either $w$ or the weighted edge density will be fixed.  
The value $w$ is assumed to be known, while parameter $\tau$ is unknown, this is the infection rate that we want to estimate based on the observed data.

\begin{figure}
\begin{minipage}{0.5\textwidth}    
    \centering\includegraphics[scale=0.5]{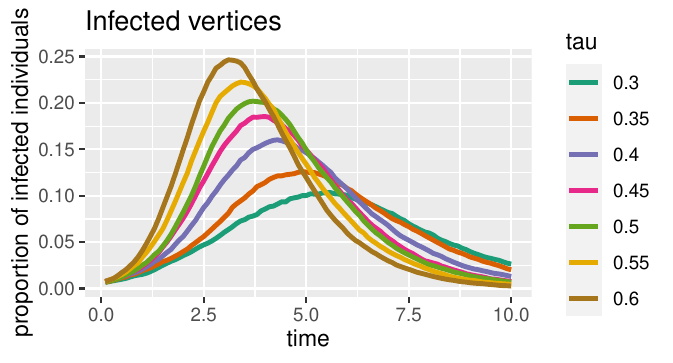}\\
    (a)
\end{minipage}
\begin{minipage}{0.5\textwidth}
\centering
\includegraphics[scale=0.5]{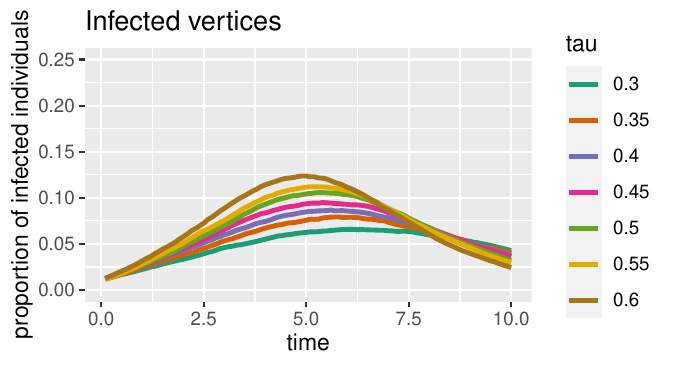}\\
(b)
    \end{minipage}
\caption{Proportion of infected vertices for $n=5000$ with (a) scale-free graph as second layer ($p_{\rm u}=0.7$, $p_{\rm tr}=0.3$, $p_{\rm pa}=0$; each curve is the average of 8 trajectories)
     and (b) with the cliques of fixed size $9$ as second layer (each curve is the average of 35 trajectories)}
    \label{fig:infecteduj}
\end{figure}

On Figure \ref{fig:infecteduj}, we see the evolution of the epidemic spread process for the household graph with scale-free second layer  and with the cliques as the second layer. In both cases, with a fixed choice of the graph parameters, for different values of $\tau=0.3, 0.35, 0.4, \ldots, 0.6$, we simulated the SIR process independently several times (with $1\%$ infected vertices at the beginning), and plotted the average of the proportion of infected vertices, as a function of time. The size of the graph ($n=5000$), the size of the households ($N_{\rm hh}=5$) and the average degree of a vertex outside its household ($8$, corresponding to $m=4$ and $N_{\rm wp}=9)$ were the same in the two cases. The size of the initial configuration of the scale-free graphs is $n_0=50$. As  expected, the spread of the epidemic is significantly more intensive in the scale-free case: vertices with very large degree have a higher chance to get infected quickly and to transmit the disease to many vertices in their neighborhood. 

\subsection{Classical estimation methods} 

In this section we briefly present a few ideas from classical methods, which are based on the maximum likelihood estimate for analoguos models and were already examined in our previous work \cite{elsocikk}. This will be useful for the analysis of the machine learning methods as well: first, we can get an idea about the epidemic statistics that  contain essential information, hence we can decide which of them should be included in the algorithms. On the other hand, we will be able to compare the results of the classical methods with the results of the machine learning estimates.

Based on the maximum likelihood estimate presented for a two-layer graph (with a complete graph as the second layer) e.g.\ in \cite{focus}, we can start from  the following formula for the estimate of the infection parameter $\tau$ at time $T$: 
  \begin{equation}\label{eq:tauhat}\hat \tau=\frac{z_{\rm I}}{\int_0^T W_t^{\rm SI}\, dt}=\frac{z_{\rm I}}{\sum_{t_i<T} W_{t_i}^{\rm SI}(t_i-t_{i-1})},\end{equation}
    where $z_{\rm I}$ is the total number of events when a vertex gets infected (until $T$), while 
$W_t^{\rm SI}$ is total weight of SI edges (edges with one susceptible and one infected endpoint) at time $t$. The times $t_i$ are the instances when an event (infection or recovery) takes place.

However, as we do not want to assume that the number (or weight) of SI edges is known exactly, we considered the estimation of this quantity, based on statistics which are probably easier to observe. In particular, we assumed that the number of within-household SI edges (denoted by $E_t^{\rm SI, hh}$) is known exactly, while we estimated the number of SI edges whose two endpoints are in different households (denoted by $E_t^{\rm SI, o}$) by
\begin{equation}\label{eq:etsio}
\hat E^{\rm SI,o}_t=I_t\cdot \bigg(d-\frac{wd}{wd+N_{\rm hh}-1}\bigg)\cdot \frac{S_t}{n}.\end{equation}
Here $d$ is the average number of the neighbors of a vertex outside its household, and $w$ is the weight of the edges going between different households.  The formula is based on a simple estimation of the probability that an infected vertex gets the disease from its own household; namely, we assume that this probability is given by the proportion of the total weight of edges inside the household and altogether, and we also replace the degree of the vertex with the average degree.
Thus the estimate for $W_t^{\rm SI}$ is
$\hat{W}_t^{\rm SI} = E_t^{\rm SI, hh} + w\cdot 
\hat E^{\rm SI,o}_t$.

It is also worth examining a variant of formula \eqref{eq:etsio} for the inhomogenous, polynomial graph model. In the formula $d$ represents the average degree all over the graph. Since vertices with larger degree get infected more easily, we can assume that the average degree of a uniformly randomly chosen infected vertex is different from $d$ and is varying over time: it is larger than $d$ in the first, increasing period of the epidemic spread, and it is smaller than $d$  in the second, "calmer" period. Hence we also consider a modified version of the above formula: 
\begin{equation}\label{eq:etilde}\tilde E^{\rm SI,o}_t=I_t\cdot \bigg(d_t^{\rm I}-\frac{wd_t^{\rm I}}{wd_t^{\rm I}+N_{\rm hh}-1}\bigg)\cdot \frac{S_t}{n},\end{equation}
where $d_t^{\rm I}$ is the average degree of the infected vertices at time $t$, counting only the neighbors of the vertex which are in a different household.  

\begin{figure}
    \centering\includegraphics[scale=0.5]{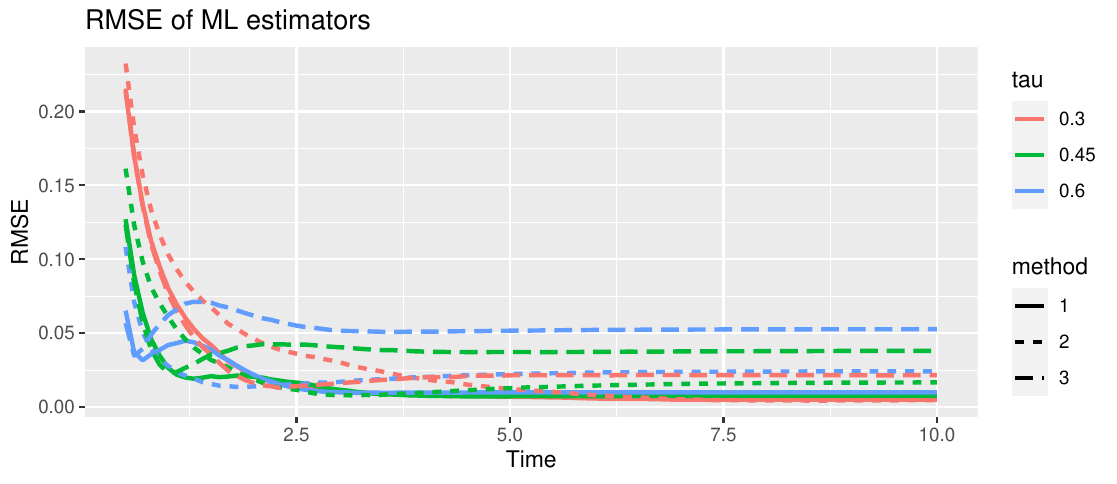}
\caption{ Root mean square error in the two-layer graph with scale-free graph as the second layer, as a function of time, according to estimates \eqref{eq:tauhat}, \eqref{eq:etsio} and \eqref{eq:etilde}. ($n=5000$ vertices) } 
\label{fig:rmse-classic}
\end{figure}

We evaluate the goodness of the estimates  by using the root mean square error, which is the square root of the average of $(\hat {\tau}-\tau)^2$, averaged for all scenarios under consideration (e.g. graph parameters, time instances, infection parameters, simulated epidemic trajectories). The simulation results based on equations \eqref{eq:tauhat}, \eqref{eq:etsio} and \eqref{eq:etilde} can be seen on Figure \ref{fig:rmse-classic} for the scale-free second-layer graph.  
Observe that the estimate $\tilde E^{\rm SI,o}_t$ (method 3) leads to somewhat better results in the first period of the epidemic spread, hence it might be worth including this quantity in the training set of the machine learning methods as well.  

\subsection{Learning the infection parameter with XGBoost}

XGBoost (eXtreme Gradient Boosting) is a powerful algorithm, where weak learners (decision trees) are combined iteratively in order to get a strong method. It is suitable for regression type problems, like ours -- but it is not especially designed for time series. As it is rather quick for the moderate data sizes we have, we were able to overcome this problem by running from scratch for each time point $T$, based on observed data $X_1,X_2,\ldots,X_T$. 

Since for the XGBoost algorithm, we need data instances of the same dimension, we recorded observations about the epidemic at equidistant time points, mimicking daily reports.
For each simulated trajectory, we registered the following quantities at the time points $t$ of the grid:
(i) $S_t, I_t, R_t$, the number of susceptible, infected and recovered vertices, respectively,
(ii) $E_t^{\rm SI,o}$ and $E_t^{\rm SI, hh}$, the number of $\rm SI$ edges outside and within households, respectively, and (iii) $d^{\rm S, w}_t$ and $d^{\rm I, w}_t$, the average weighted degree of susceptible resp. infected vertices.
Since the quantities (iii) are constant in the graph model where the second layer  consists of fixed size cliques, in that case we did not use them.

The first three quantities arise naturally. It is worth mentioning that the number of recovered vertices at time $t$ is $R_t=n-S_t-I_t$, so $R_t$ does not give new information, since $n$ is the same for all graphs in the training and test set. The use of $E^{\rm SI, o}_t$ and $E^{\rm SI, hh}_t$ is based on equation \eqref{eq:tauhat}; as we have seen, this gives a very good estimate by using the exact number of $\rm SI$ edges. Similarly, equation \eqref{eq:etilde} suggests that it might be worth involving $d^{\rm I}_t$. 
 As for  $d^{\rm S}_t$ (that is, the average degree of the susceptible vertices),  this quantity might also  contain information on the actual state of the epidemic: we expect that this quantity decreases, and if this has a relatively large value, then there are still vertices with larger degree that have a high chance to be infected, so we are not at the end of the epidemic process.  

It is important to emphasize that neither the graph, nor the exact time of the infection or recovery events are used for the XGBoost parameter learning, the only available data is the time series of the above variables, 
observed at the equidistant time points.
In the chosen parametrisation this was achieved by using $0.1$ units as the time step, so the values of $t$ were $0.1, 0.2, \ldots, 29.9, 30$. Recall that the time unit was chosen by fixing the recovery rate $\gamma = 1$. When we run the algorithm for a time $T$,
we use all the observed data up to time $T$. 

We note that in some experiments, instead of all variables above, we used only (i).
The results when using all information (i)-(iii) were compared with the results of the classical estimate \eqref{eq:tauhat} with $W_t^{\rm SI}$ known exactly. On the other hand, the results when using only information (i) were compared with the results of the classical estimate where the number of $\rm SI$ edges is estimated by \eqref{eq:etsio} (see  Tables 1 and 2 in the Results section). 

We used the pre-programmed function in the R package 'xgboost', with booster gbtree (gradient boosted tree). Most parameters were set at their default values, but we modified those where our grid search suggested an improvement of performance. For the sake of completeness, these were:
eta $= 0.2$, subsample $= 0.9$, nrounds $=90$, colsample\_bytree $=0.8$ and
min\_child\_weight $ =0.7$.

\subsection{Application of neural networks}
Using artificial neural networks is becoming an increasingly prominent approach in parameter estimation and epidemic modeling.
First, we examined the generalizability of the neural network method. We have found that in case of one variable graph model parameter, the neural network performs adequately, and comparably with the classical method. Upon adding multiple variables, the neural network loses accuracy significantly.

Our chosen model was a convolutional neural network (CNN), which is generally used to process spatially correlated data (in our case the ordered structure is provided by the temporal process). CNN structure is biologically motivated and relies on feature extraction, increasing the number of feature channels with each additional neuron layer in the network. Our model follows a similar design principle with 1-dimensional convolutional layers. Each layer has different convolutional kernels of size 4. The number of kernels in each layer is the following: $N_{input}$, 16, 64, 256, 512, 1024, 128, 1, where $N_{input}$ is the number of time series used in the setting (either only S, I, R time series are used or also the time series for the number of SI edges). Additionally, there were 2 fully connected layers after the convolutional layers.

The activation functions were leaky rectified linear unit functions to avoid the so-called “dying ReLU problem”. We used Adam optimization algorithm, which is considered the standard state-of-the-art choice. The learning rate was chosen to be 0.00002 and the slope on the leaky domain was 0.2, based on a hyperparameter-optimizing grid search.

The model was implemented using PyTorch \cite{paszke2019pytorch}.

\begin{figure}
    \centering\includegraphics[scale=0.5]{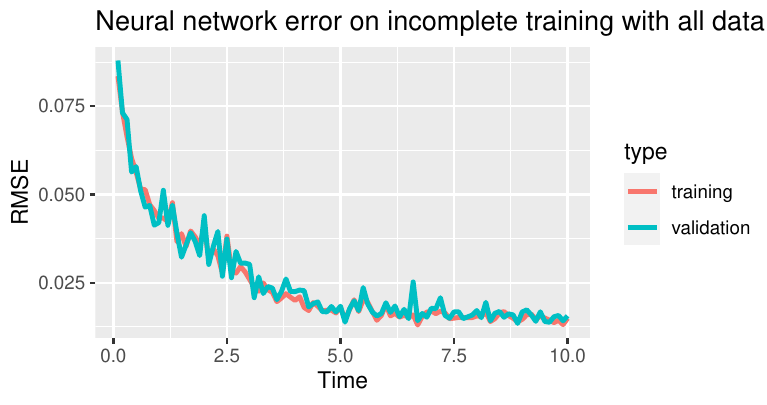}
      \caption{The performance of the CNN architecture improves as measured by RMSE loss when trained on increasing amount of information about the development of the epidemic spread in time. In this setting information about the number of SI edges is known. The $\tau$ parameter is in the range $[0.3,0.6]$. The performance is validated on a sample of 260 time series for each datapoint.}
    \label{fig:cnn_all_data}
\end{figure}

We explored the question of whether it is possible to have a single viable predictor applicable to all stages of the epidemic, or if training a distinct CNN model for all initial segment lengths always gets a better result overall.

First, we evaluated a model trained only on time series that contain the whole process of the epidemic to its completion and determined that upon receiveing shorter initial segments of the epidemic, the model behaves in a less accurate and rather unpredictable way, as expected. Training on a larger dataset of the same type did not affect the results.

Next, we used a training dataset which contains initial segments of different sizes, with the lengths uniformly distributed. The results were promising as the error rate was much smaller for a large portion of the epidemic (until time unit 7), although surprisingly after this point in time the prediction becomes less accurate. The model is much more predictable as the curve of the error rate is significantly smoother.

Considering the concavity of the error rate in the previous case it seemed that the CNN model optimizes the prediction for segment lengths in the middle range. Thus, we suspected that the initial segments of such lengths might be overrepresented in the dataset. We employed sampling strategies to address this issue. Instead of using a uniform distribution for segment lengths, we used a beta distribution with parameters $\alpha = \beta = 0.5$ which gives higher representation to full length time series and very short ones. The results confirmed the suspicion, since the error curve lost its concavity and the prediction is minimal at full length. However, the prediction barely gets any better after time unit 4. The performance did not improve upon increasing the dataset.

Upon skewing the beta distribution to favour full length segments over very short ones, the performance improved sligthly, although it still fell short of the performance of the first training strategy at the end of the epidemic.

\begin{figure}[h]
  
\centering
 \includegraphics[width=0.5\textwidth]{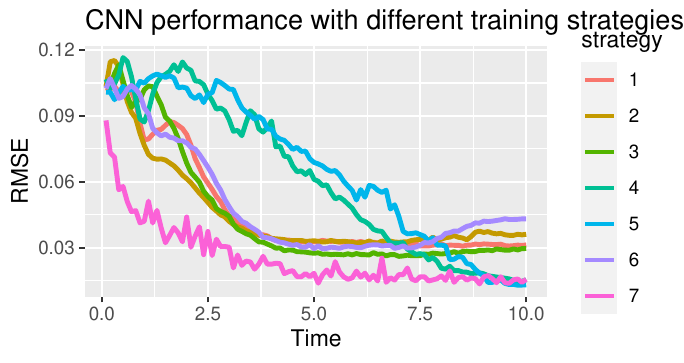}
    \caption{Comparing the performance of the CNN model as a function of the progression of the epidemic using different training strategies: 1 - training datapoints sampled with symmetric beta distribution; 2 - same sampling as 1, but larger dataset; 3 - sampling distribution is a beta distribution skewed towards larger values, to include more datapoints with longer time series; 4 - training dataset containing exclusively full-length time series; 5 - same dataset as 4 with more datapoints; 6 - training datapoints sampled with uniform distribution; 7 - for each timepoint a separate model was trained with initial segments of time series ending at each given timepoint. Only S,I,R data were used and the underlying network model had a second layer generated by the polynomial model.}
    \label{fig:cnn_compare}
\end{figure}

The summmary of these results can be seen in figure (\ref{fig:cnn_compare}). Neither of the strategies above resulted in better performance than training distinct CNN models on each time segment length individually. However, using a single model and training it on different segment lengths improved the consistency of the predictor and has the obvious advantage of using only a single model, which has to be trained only once.

\section{Results}
\label{sec:results}
\begin{figure}
\begin{minipage}{0.47\textwidth}    
  
\centering
    \includegraphics[width=1\textwidth]{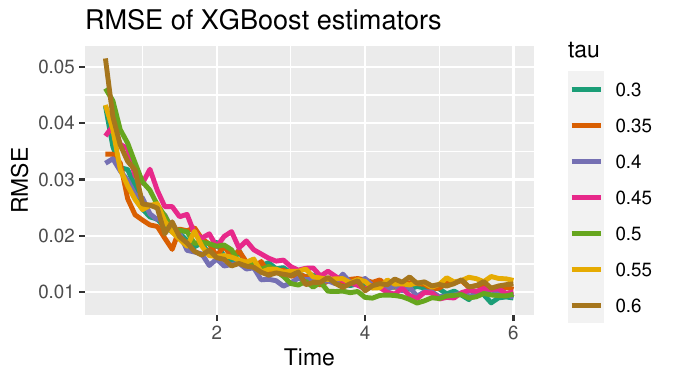}
    \end{minipage}
\begin{minipage}{0.53\textwidth}
\centering
 \includegraphics[width=1\textwidth]{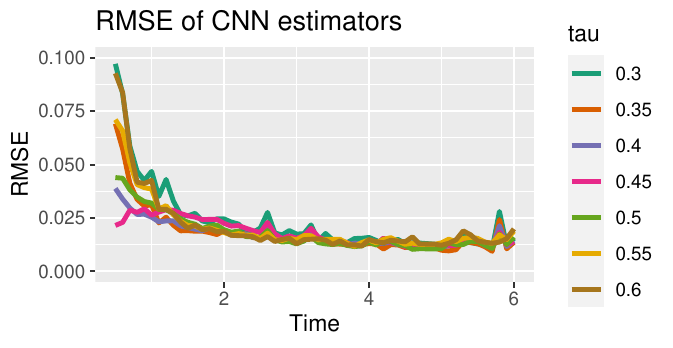}
     \end{minipage}
   \caption{Left panel: The RMSE as the function of time for different $\tau$ values. $N=5000$ with relaxed complete cliques as second layer. Each curve is the average of $90$ trajectories.  Right panel: Comparing the performance of the CNN model as a function of the progression of the epidemic for different values of $\tau$. Only S,I,R data were used and the underlying network model had a second layer of cliques.}
    \label{fig:xgb1}
\end{figure}

\begin{figure}
\begin{minipage}{0.5\textwidth}    
  
\centering
 \includegraphics[width=1\textwidth] {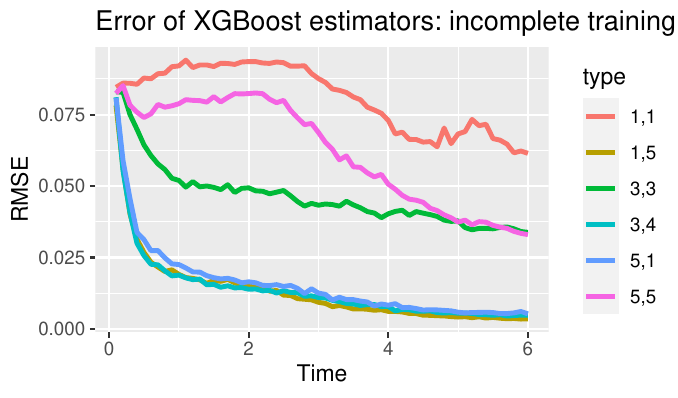}
     \end{minipage}
     \begin{minipage}{0.5\textwidth}    
\centering
 \includegraphics[width=1\textwidth]{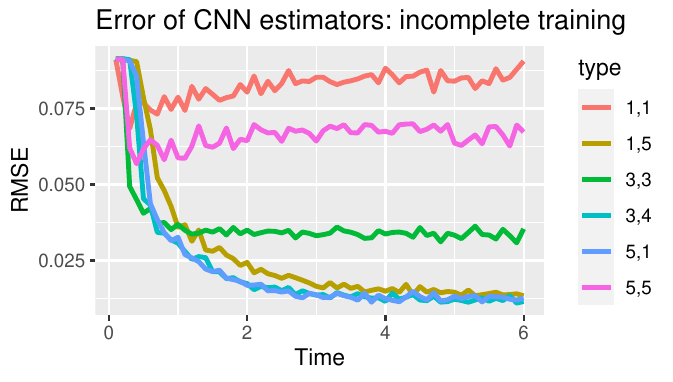}
     \end{minipage}
     \caption{The RMSE as the function of time for different training and test sets. $N=5000$ with full cliques as second layer. Left panel: XGBoost results. Each curve is the average of $465$ trajectories. Right panel: CNN results.}
    \label{fig:xgb2}
\end{figure}

The figures  \ref{fig:cnn_all_data} and \ref{fig:xgb1} are  typical results for the out-of-sample estimations in case of using all the above information and simulations for all available parameter values. It is worth noting that the  machine learning methods 
performed better for early timepoints quicker  than the parametric estimators, but their error rates stabilised after 5 time units, not showing further improvements.
We have also investigated the effect of missing configurations from the training. The results are interesting -- they very much depend on the structure of the underlying graphs. In case of different average degrees, the effect is substantial. The left panel of Figure \ref{fig:xgb2} shows different cases, labeled as follows: the first label shows the omitted index from the sequence of potential clique sizes from the set $\{7,8,9,10,11\}$, while the second one refers to the index of the clique size in the test set. 

  

\begin{figure}
\begin{minipage}{0.5\textwidth}    
  \centering
 \includegraphics[width=1\textwidth]{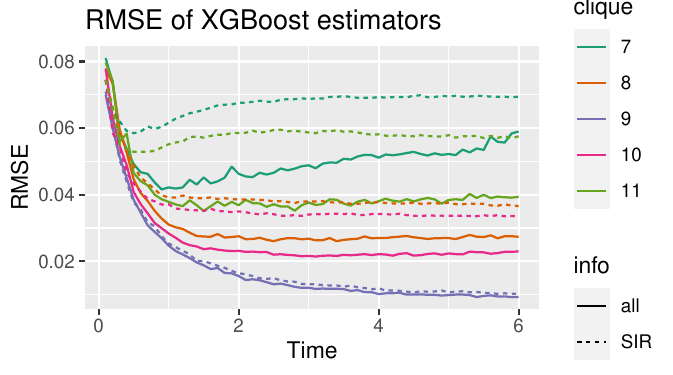}
     \end{minipage}
\begin{minipage}{0.5\textwidth}    

 \includegraphics[width=1\textwidth]{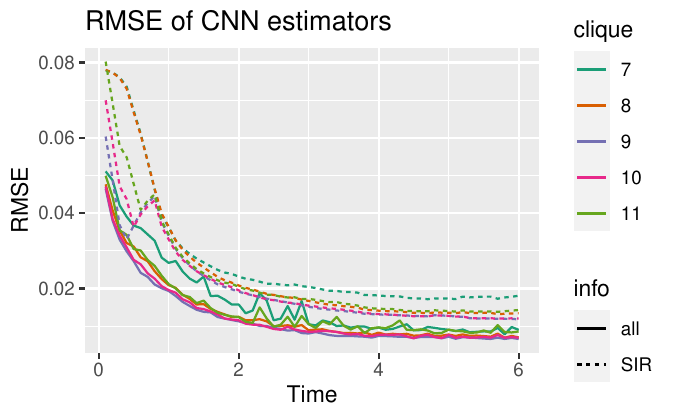}
   \end{minipage}
     \caption{ The RMSE as the function of time for different test sets  and learning information. $N=5000$ with full cliques as second layer. The plots show the averages over $\tau$. Left panel: XGBoost results. Each curve is the average of $465$ trajectories. Right panel: CNN results. Each curve is the average of $4650$ trajectories. 
     }
    \label{fig:fw2}
  \end{figure}

   
In the next experiment we have ensured a fixed density of the graph for different clique sizes by changing the value of the weight $w$. The baseline was the case of 9-sized cliques and $w=0.4$, and for the clique-size $7,8,\ldots,11$ the weights were adjusted. When training was carried out by these graphs and in the test the clique size was 9, combined with the customary $w=0.4$, 
then the RMSE values varied according to the 
clique size: the more different the graph in the testing set, the worse the estimation was (especially for XGBoost, see the left panel of Figure \ref{fig:fw2}). However, the CNN model seemed to be much less sensitive to different test datasets if the graph density was constant. Here we have also investigated the importance of the other information beyond the number of SI(R) vertices. It is an interesting observation in case of the XGBoost that it is more important if the testing is carried out for a graph with different clique size (e.g. 7 or 11) and the effect fades away for the case when both have the same density (clique size 9). It has to be noted that for the CNN to achieve the shown very good results, one has to carefully tune the parameters of the network.

Finally, we compared the performance (measured by the RMSE) of the considered estimation methods for two scenarios: (i)
for two-layer graphs with a polynomial graph as second layer and (ii) for two-layer graphs with full cliques with fixed edge density as second layer, see Tables \ref{politabla} and \ref{klikktabla}, respectively. For both scenarios, the infection parameter $\tau$ had $31$ possible values, ranging from $0.3$ to $0.6$ in steps of $0.01$.

The test data was the same for all methods, moreover, the XGBoost and CNN methods used the same training dataset, disjoint from the test data (the classical ML method does not use training data). $70\%$ of all data were assigned as training data, and $30\%$ as test data. For both scenarios, we calculated the overall RMSE for fixed time instances of the epidemic.

For scenario (i), the polynomial second layer, the training set contained $10850 = 10 \times 31 \times 35$ epidemics ((number of polynomial graph parameter triplets) $\times$ (number of $\tau$ values) $\times$ (number of replications for each combination)). The test set contained $4650 = 10 \times 31 \times 15$ epidemics. The graph parameter triplets were as follows: In the first 5 cases, $p_{\rm pa} =0$, while $p_{\rm tr}$ took values $0.1, 0.15, 0.2, 0.25, 0.3$. In the last 5 cases, on the other hand, $p_{\rm tr} =0$, while $p_{\rm pa}$ took values $0.1, 0.15, 0.2, 0.25, 0.3$. The upper half of Table \ref{politabla} summarises the results for methods where detailed information was available (in particular, the number of SI edges), while the lower half contains methods where only the S/I/R counts were used for estimation.

\begin{table}
\caption{Overall RMSE of the estimate of $\tau$ for various methods, for two-layer graphs with polynomial second layer. The columns show the elapsed time from the start of the epidemic. 
}
\label{politabla}
\begin{tabular}{l|c|c|c|c|c|}
method & $t=1$ & $t=2$ & $t=4$ & $t=6$ & $t=10$ \\
\hline
ML, $E^{\rm SI, o}$ known & 0.0440 & 0.0204 & {\bf 0.0085} & {\bf 0.0077} & {\bf 0.0077}\\
\hline
XGBoost, all information & {\bf 0.0314} & {\bf 0.0167} & 0.0104 & 0.0088 & 0.0084 \\
\hline
CNN, all information& 0.0365 & 0.0241 & 0.0193 & 0.0147 & 0.0121\\
\hline
\hline
ML, $E^{\rm SI, o}$ estimated by \eqref{eq:etsio}& 0.0654 & 0.0261 & 0.0142 & 0.0156 & 0.0171\\
\hline
ML, $E^{\rm SI, o}$ estimated by \eqref{eq:etilde}. & 0.0457 & 0.0428 & 0.0380  & 0.0382 & 0.0387\\
\hline
XGBoost, only SIR & {\bf 0.0390} & {\bf 0.0206} & {\bf 0.0117} & {\bf 0.0100} & {\bf 0.0095} \\
\hline
CNN, only SIR & 0.0450 & 0.0281 & 0.0151 & 0.0134 & 0.0121 \\
\hline
\end{tabular}
\label{tab:all_methods_dec}
\end{table}

For scenario (ii), the clique second layer, 
the training set contained $32550 = 6 \times 31 \times 175$ epidemics  ((number of caveman graph parameters) $\times$ (number of $\tau$ values) $\times$ (number of replications for each combination)). The test set contained $13950 = 6 \times 31 \times 75$ epidemics (since the simulation of the process is much faster in this case than in the polynomial model, we had the possibility to generate larger data sets). As for the parameters of the second layer, we used six different clique sizes $N_{\rm wp} = 7, 8, 10, 11, 12, 15$. The corresponding between-household edge weight $w$ was chosen such that $(N_{\rm wp}-1) \times w = 3.2$, i.e., the edge density is constant.
The structure of Table \ref{klikktabla} is the same as before, only the estimation of $E^{\rm SI,o}$ by equation \eqref{eq:etilde} was not applicable here. 

Looking at the tables, we see very similar patterns for the two scenarios.
In the early phase of the epidemics ($t=1$), the XGBoost and CNN methods perform better than the classical ones. Moreover, the error of the machine learning methods is not much worse when they use only the SIR counts. At the same time, the estimation of  $E^{\rm SI, o}$ by \eqref{eq:etilde} seems to be more precise in this initial stage of the epidemics.  Later on, however, the estimate \eqref{eq:etilde} breaks down. In the still rising phase of the epidemics ($t=2$), the methods give comparable results, but the XGBoost and CNN methods are slightly better than the classical ones. Later on ($t\geq 4$) the ML method with known number of SI edges is the most precise, however, the XGBoost and CNN methods using only SIR counts are not much worse -- and they are definitely better than the ML method with estimated number of SI edges. Surprisingly, the results do not improve significantly after $t=6$. Comparing the two tables, there does not seem to be a large difference between the RMSE values for the two graph models.

\begin{table}
\caption{Overall RMSE of the estimate of $\tau$ for various methods, for two-layer graphs with full cliques second layer, fixed edge density. The columns show the elapsed time from the start of the epidemic.}
\label{klikktabla}
\begin{tabular}{l|c|c|c|c|c|}
method & $t=1$ & $t=2$ & $t=4$ & $t=6$ & $t=10$ \\
\hline
ML, $E^{\rm SI, o}$ known & 0.0515 & 0.0195 & \bf{0.0093} & \bf{0.0087} & \bf{0.0087} \\
\hline
XGBoost, all information & \bf{0.0270} & 0.0160 & 0.0110 & 0.0100 & 0.0090 \\
\hline
CNN, all information& 0.0295 & \bf{0.0152} & 0.0109 & 0.0105 & 0.0103 \\
\hline
\hline
ML, $E^{\rm SI, o}$ estimated by \eqref{eq:etsio}& 0.0441 & 0.0419 & 0.0303 & 0.0255 & 0.0236 \\
\hline
XGBoost, only SIR & \bf{0.0280} & \bf{0.0190} & \bf{0.0140} & \bf{0.0130} & \bf{0.0120} \\
\hline
CNN, only SIR & 0.0303 & 0.0208 & 0.0162 & 0.0148 & 0.0151 \\
\hline
\end{tabular}
\label{tab:all_methods2}
\end{table}

\section{Conclusions}

We have explored several methods for estimating the infection rate of a simulated SIR process on different graph structures using aggregated data about the states of the nodes and infectuous contacts. In all cases, the estimate increases in accuracy as the known initial segment of the data time series becomes increasingly longer. When all available types of information are used, the classical method based on maximum likelihood estimation performs best for longer time segments. However, if data reflecting the graph structure is missing, both machine learning algorithms outperform the classical metod. On the other hand, the methods performed comparably for both investigated graph models.

The XGBoost algorithm proved to be best overall, although it seems less able to generalize in order to assess data that is very different from the training dataset. The convolutional neural network is seemingly less sensitive to the training dataset, i.e. when the weighted edge-density of the graph underlying the SIR process is available during the training process, the CNN provides relatively accurate estimates even for otherwise dissimilar data. However, the CNN model is computationally more demanding than the other methods and by reducing this demand, the method loses accuracy significantly.

\bibliography{xgboostcnn}
\bibliographystyle{plain}

\end{document}